\documentclass[12pt]{article}
\usepackage{verbatim,a4wide,epsfig,amsmath,url,amssymb,xcolor}

\usepackage[round]{natbib}

\title{One-Sided Cross-Validation for Nonsmooth Density Functions}
\author{Olga Y. Savchuk}
\date{}

\begin{document}

\maketitle

\vspace{-1.2cm}
\begin{abstract}
\noindent One-sided cross-validation (OSCV) is a bandwidth selection method initially introduced by~\citet{HartYi} in the context of smooth regression functions.
\citet{OSCV:density} developed a version of OSCV for smooth density functions. This article extends the method for nonsmooth densities. It also introduces the fully robust OSCV modification that produces consistent OSCV bandwidths for both smooth and nonsmooth cases. Practical implementations of the OSCV method for smooth and nonsmooth densities are discussed. One of the considered cross-validation kernels has potential for improving the OSCV method's implementation in the regression context. 
\end{abstract}

\section{Introduction}

Let $X_1,\,X_2,\ldots,X_n$ be a random sample from the probability density function $f$. The kernel density estimator of $f$ is computed as
\[
\hat f_{h,K}(x)=\frac{1}{nh}\sum_{i=1}^nK\left(\frac{x-X_i}{h}\right),
\]
where $h>0$ is a smoothing parameter that is usually called the bandwidth, and the kernel $K$ is assumed to be of the second order, which means that $\int_{-\infty}^\infty K(u)\,du=1$, $\int_{-\infty}^\infty uK(u)\,du=0$, and $\int_{-\infty}^\infty u^2K(u)\,du<\infty$. Most frequent choices of $K$ include the Gaussian, Epanechnikov, and quartic kernels (see~\citet{Wand:KS}). 

The two most commonly used measures of performance of $\hat f_{h,K}$ are the integrated squared error (ISE) and the mean integrated squared error (MISE) defined as
\[
\begin{array}{l}
\displaystyle{\mbox{ISE}_K(h)=\int_{-\infty}^\infty \left(\hat f_{h,K}(x)-f(x)\right)^2\,dx,}\\
\displaystyle{\mbox{MISE}_K(h)=\mbox{E}\left(\mbox{ISE}_K(h)\right)}.
\end{array}
\]
Let $\hat h_0$ and $h_0$ denote the minimizers of the $\mbox{ISE}_K$ and $\mbox{MISE}_K$ functions, respectively.

Both bandwidths $\hat h_0$ and $h_0$ are unavailable for practical use since their computation requires knowing $f$. There exist many data-driven bandwidth selection techniques (see the survey of~\citet{JonesMarronSheather:survey}). Some of the most popular bandwidth selectors are the plug-in rule of~\citet{Sheather:PI} and the least squares cross-validation (LSCV) method proposed independently by~\citet{Rudemo:LSCV} and~\citet{Bowman:LSCV}.

The cross-validation method is quite popular among practitioners because of its simplicity. Moreover, it requires fewer assumptions on $f$ compared to the plug-in method. Nevertheless, the method is criticized because of producing too variable bandwidths and selecting the trivial bandwidths for the data sets that contain substantial amount of tied observations (see~\citet{Silverman:book} and~\citet{Chiu:rounding}). A well known LSCV paradox consists in the method's improved performance on the harder estimation problems (see~\citet{Loader:Classical}). A couple of successful modifications of the LSCV method that take advantage of this paradox are the one-sided cross-validation (OSCV) method  proposed by~\citet{OSCV:density} and the indirect cross-validation (ICV) method of~\citet{SavchukHartSheather:ICV}. Both methods are supported by the corresponding R packages (see~\citet{ICV:package} and~\citet{OSCV:package}).

The OSCV method is originally introduced in the regression context (see~\citet{HartYi}). ~\citet{OSCV:density} extended the method to the case of smooth densities. A density function $f$ is referred to as {\it smooth} if it is twice continuously differentiable, whereas it is called {\it nonsmooth} if it is continuous but has finitely many simple discontinuity points in its first derivative. The OSCV method in the smooth case is shown to greatly stabilize the bandwidth distribution (see~\citet{OSCV:density}). This article extends the OSCV method to the case of nonsmooth density functions.

\citet{OSCV:density} introduced the left-sided and right-sided OSCV versions based on the so-called left-sided and right-sided kernels, respectively. Both one-sided kernels are obtained by multiplying a benchmark two-sided kernel by a linear function and restricting the support of the one-sided kernel to either $(-\infty,0]$ (left-sided case) or $[0,\infty)$ (right-sided case). 
In this article we restrict our attention on the right-sided OSCV version.

The right-sided kernel $L$ based on the benchmark two-sided kernel $H$ is computed as
\begin{equation}
\label{eq:L}
\displaystyle{L(u)=\frac{\int_0^\infty t^2H(t)\,dt-u\int_0^\infty tH(t)\,dt}{\int_0^\infty H(t)\,dt \int_0^\infty t^2H(t)\,dt-\left(\int_0^\infty tH(t)\,dt\right)^2}\,H(u)I_{[0,\infty)}(u)}.
\end{equation}
It follows that $L$ is of the second order. Generally, the benchmark kernel $H$ is different from the kernel $K$ used to compute $\hat f_{h,K}$. \citet{OSCV:density} implemented OSCV based on $H=K=K_E$, where $K_E$ denotes the Epanechnikov kernel.

The OSCV function based on a one-sided kernel $L$ is defined as
\begin{equation}
\label{eq:OSCV}
\mbox{OSCV}_L(b)=R(\hat f_{b,L})-\frac{2}{n}\sum_{i=1}^n\hat f_{b,L}^{-i}(X_i),
\end{equation}
where the definition of $R(\cdot)$ is given in the Appendix. In the above expression $\hat f_{b,L}$ is the density estimator based on the kernel $L$ and the bandwidth $b$, whereas $\hat f_{b,L}^{-i}(X_i)$ is its leave-one-out modification that is computed from all data points except $X_i$. The above version of the OSCV function mimics the traditional definition of the CV function of~\citet{Rudemo:LSCV} and~\citet{Bowman:LSCV} and slightly differs from the function used in~\citet{OSCV:density} and the follow-up articles (see~\citet{Mammen:Do-validation} and~\citet{Mammen:ICV}). Indeed, in the OSCV function of~\citet{OSCV:density}, the estimator $\hat f_{b,L}^{-i}$ under the sum is replaced by $\hat f_{b,L}$.  This is justified by assuming $L(0)=0$. We find this assumption rather restrictive since the one-sided versions of the most frequently used kernels do not possess this property. Since the case $L(0)\neq 0$ does not substantially complicate the OSCV implementation, we proceed by using~\eqref{eq:OSCV}. Let $\hat b$ denote the minimizer of~\eqref{eq:OSCV}.

\citet{OSCV:density} defined the OSCV bandwidth in the smooth case as $\hat h_{OSCV}=C\cdot\hat b$, where
\begin{equation}
\label{eq:C}
C=\left(\frac{R(K)}{R(L)}\cdot\frac{\mu_{2}^2(L)}{\mu_2^2(K)}\right)^{1/5}.
\end{equation}
The functionals $R(\cdot)$ and $\mu_2(\cdot)$ are defined in the Appendix. The OSCV bandwidth $\hat h_{OSCV}$ is consistent for the MISE optimal bandwidth $h_0$, that is $\displaystyle{\hat h_{OSCV}\overset{p}{\to} h_0}$.

\section{OSCV for nonsmooth density functions\label{sec:OSCV_nonsmooth}}

In this section we extend the OSCV algorithm to the case of a nonsmooth density $f$ that has simple discontinuities in its first derivative at the points $\{x^{(t)}\}$, $t=1,2,\ldots,k$. The extension is based on the asymptotic expansion of MISE of the kernel density estimator in the nonsmooth case (see~\citet{Cline_Hart:nonsmooth_dens} and~\citet{vanEs}) that has the following form:
\[
\mbox{MISE}_K(h)=\mbox{AMISE}_K^*(h)+o\left(h^3+\frac{1}{nh}\right),
\]
where
\[
\mbox{AMISE}_K^*(h)=\frac{R(K)}{nh}+h^3B(K)\sum_{i=1}^k\left(f^\prime\left(x^{(t)}+\right)-f^\prime\left(x^{(t)}-\right)\right)^2,
\]
where $B(\cdot)$ is defined in the Appendix. The above expression yields the following asymptotically optimal bandwidth $h_n^*$:
\[
h_n^*=\left(\frac{R(K)}{3B(K)\sum_{i=1}^k\left(f^\prime\left(x^{(t)}+\right)-f^\prime\left(x^{(t)}-\right)\right)^2}\right)^{1/4}n^{-1/4}.
\]
The asymptotic expansion of $\mbox{MISE}_L(b)$, the MISE function for $\hat f_{b,L}$, has the same form as that of $\mbox{MISE}_K(h)$ with $K$ and $h$ being replaced by $L$ and $b$. Let $b_n^*$ denote the asymptotically optimal bandwidth for $\mbox{MISE}_L(b)$. It then follows that
\begin{equation}
\label{eq:Cstar}
\frac{h_n^*}{b_n^*}=C^*=\left(\frac{R(K)}{B(K)}\cdot
\frac{B(L)}{R(L)}\right)^{1/4}.
\end{equation}
This motivates defining the OSCV bandwidth in the nonsmooth case as $\hat h_{OSCV}=C^*\cdot\hat b$.

In both smooth and nonsmooth cases the OSCV bandwidth is defined by multiplying $\hat b$, the minimizer of the OSCV function~\eqref{eq:OSCV}, by a rescaling constant. The constant $C$~\eqref{eq:C} is used in the smooth case, whereas the constant $C^*$~\eqref{eq:Cstar} is used in the nonsmooth case.

It is remarkable that the expressions for $C$ and $C^*$ are identical to those in the OSCV implementation for regression functions (see~\citet{Savchuk:OSCVnonsmooth} and~\citet{Savchuk:Corrigendum}). This follows from similarity of the corresponding asymptotic expansions of MISE of the kernel density estimator and the mean average squared error (MASE) of the local linear estimator. The values of $C$ and $C^*$ in the case $H=K$ for the most frequently used kernels $K$ and their one-sided counterparts $L$ are given in Table~\ref{tab:Constants}.
\begin{table}
\begin{center}
\begin{tabular}{|l||c|c|c|}
\hline
Kernel $K$&$C$&$C^*$&$E_C$\\
\hline \hline
Epanechnikov&0.5371&0.5019&7.01\\
\hline
quartic&0.5573&0.5206&7.05\\
\hline
Gaussian&0.6168&0.5730&7.64\\
\hline
\end{tabular}
\caption{Rescaling constants for the most frequently used kernels.\label{tab:Constants}}
\end{center}
\end{table}
The quantity $E_C$ that appears in the last column of Table~\ref{tab:Constants} is defined by
\[
E_C=\left(\frac{C}{C^*}-1\right)\cdot100\%.
\]
$E_C$ assesses the magnitude and direction of the asymptotic relative bandwidth bias introduced by using $C$ instead of $C^*$ in the nonsmooth case.

In this article we set $K=\phi$, where $\phi$ denotes the Gaussian kernel, and assess how different one-sided kernels change the theoretical properties and practical performance of the OSCV method. In the example's section below we investigate performances of $L_G$ and $L_E$, the one-sided Gaussian and Epanechnikov kernels, respectively, for smooth and nonsmooth densities. The one-sided kernels discussed in Section~\ref{sec:FROSCV} increase the method's resistance against potential nonsmoothness of $f$, at least theoretically.

\section{Examples}

\noindent{\bf Example 1.} This example shows that in the nonsmooth case the actual bias of $\hat h_{OSCV}$ based on the smooth constant $C$ tends to be even greater than the corresponding value of $E_C$ from Table~\ref{tab:Constants}.

\begin{figure}[h]
\begin{center}
\begin{tabular}{cc}
{\bf(a)}&{\bf(b)}\\
\vspace{-0.5cm}&\vspace{-0.5cm}\\
\epsfig{file=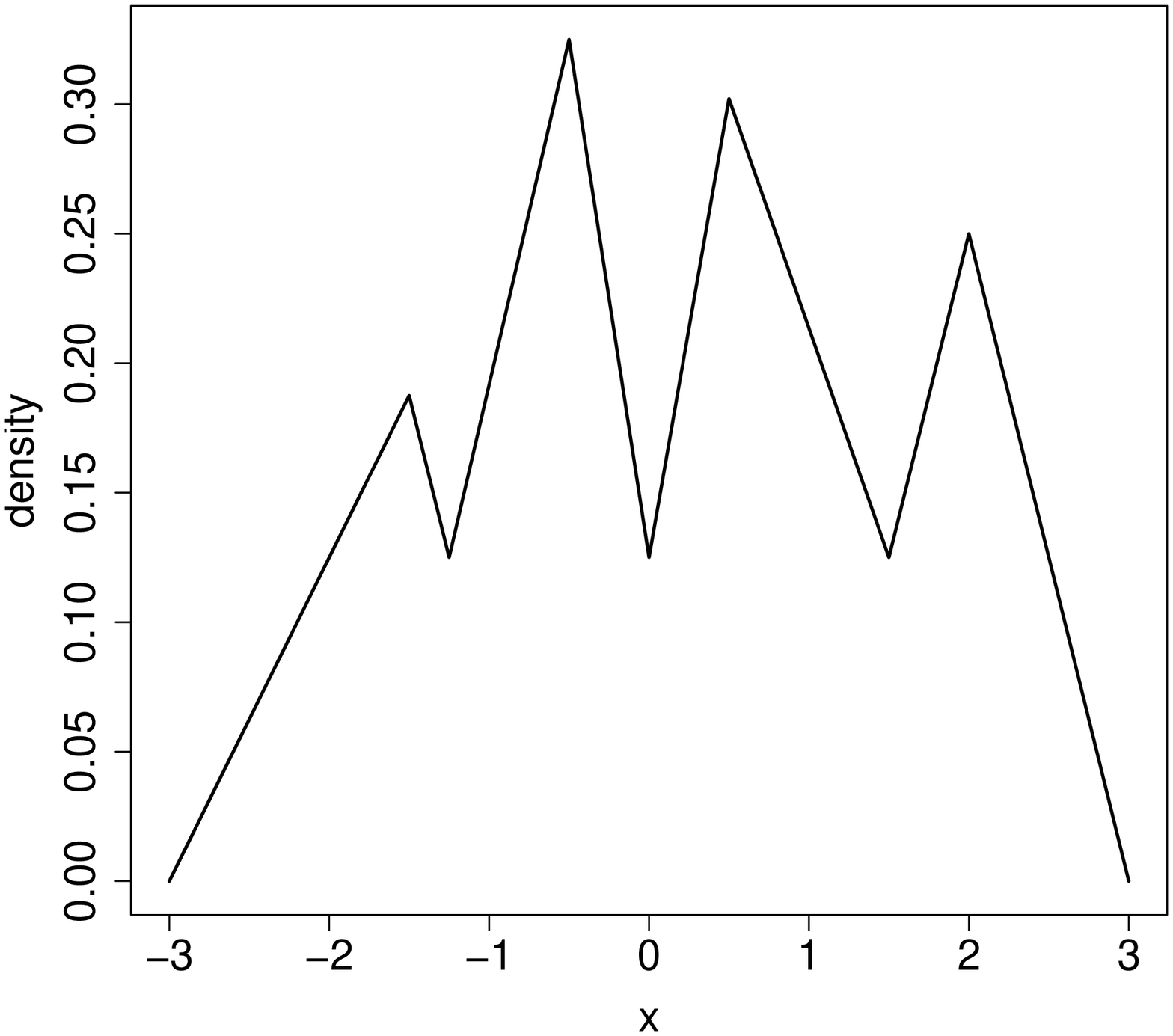,height=200pt}&\epsfig{file=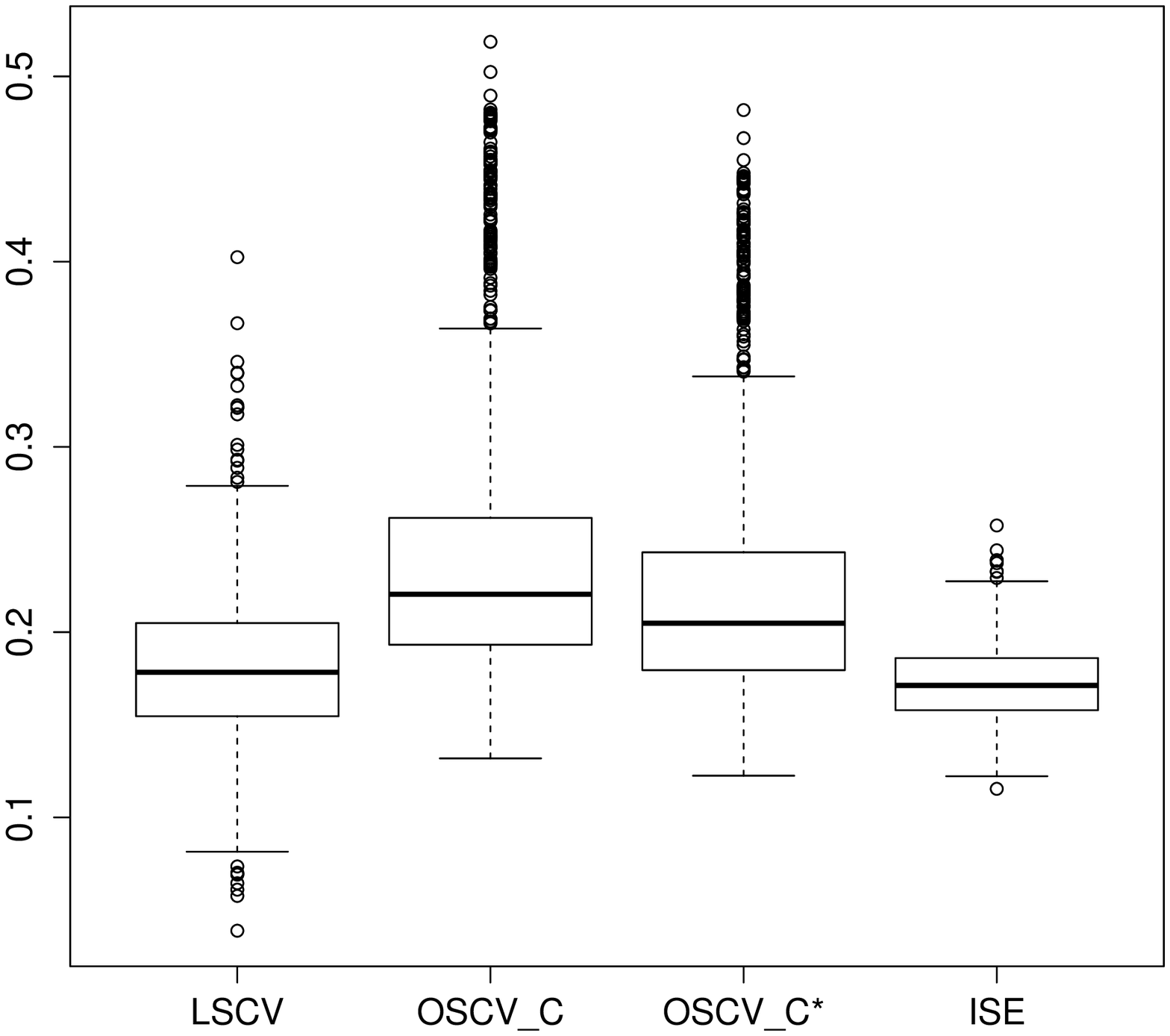,height=200pt}\\
\end{tabular}

\vspace{-0.5cm}
\caption{{\bf(a)} The density $f^*$ with 7 cusps. {\bf(b)} Boxplots for LSCV, ISE and OSCV versions based on $L_G$ that use $C$ and $C^*$.\label{fig:f_nonsmooth}}
\end{center}
\end{figure}

We generated 1000 data sets of size $n=500$ from the density $f^*$ shown in Figure~\ref{fig:f_nonsmooth}~{\bf(a)}. This density can be found in the R package \verb"OSCV" that allows generating random samples from this density and computing the $f^*$-based ISE functions (see~\citet{OSCV:package}). 

Figure~\ref{fig:f_nonsmooth}~{\bf(b)} displays the boxplots of the bandwidths selected by different methods.  Two boxplots in the middle show the distributions of OSCV bandwidths selected by $L_G$ and based on the rescaling constants $C$ and $C^*$ from Table~\ref{tab:Constants}. Also shown are the boxplots of the LSCV bandwidths and the ISE-optimal bandwidths $\hat h_0$.

We used the following measures of performance of a data-driven bandwidth $\hat h$:
\begin{equation}
\label{eq:delta_measures}
\begin{array}{l}
\displaystyle{\Delta_B=\frac{\hat M(\hat h)-\hat M(\hat h_0)}{\hat M(\hat h_0)}\cdot100\%,}\\[0.35cm]
\displaystyle{\Delta_{ISE}=\hat M\left(\frac{\mbox{ISE}(\hat h)-\mbox{ISE}(\hat h_0)}{\mbox{ISE}(\hat h_0)}\right)\cdot 100\%,}
\end{array}
\end{equation}
where $\hat M(Y)$ denotes the median of a random variable $Y$ computed over 1000 replications. In the nonsmooth case, the value of $\Delta_{ISE}$ is an empirical analog of $E_C$ for $\hat h_{OSCV}$ based on the smooth constant $C$.

\begin{figure}[h]
\begin{center}
\begin{tabular}{cc}
{\bf(a)}&{\bf(b)}\\
\vspace{-0.75cm}&\vspace{-0.75cm}\\
\epsfig{file=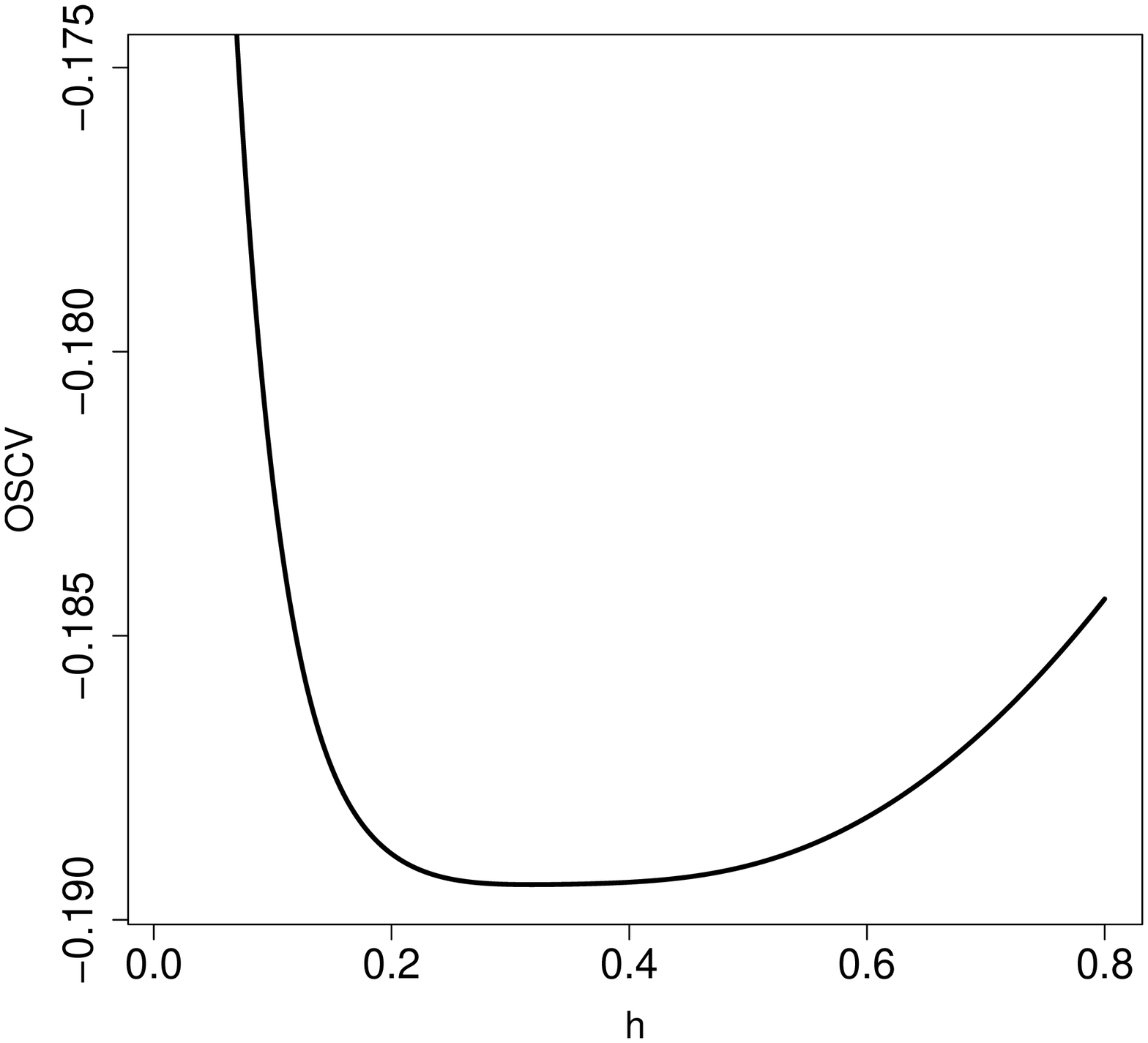,height=200pt}&\epsfig{file=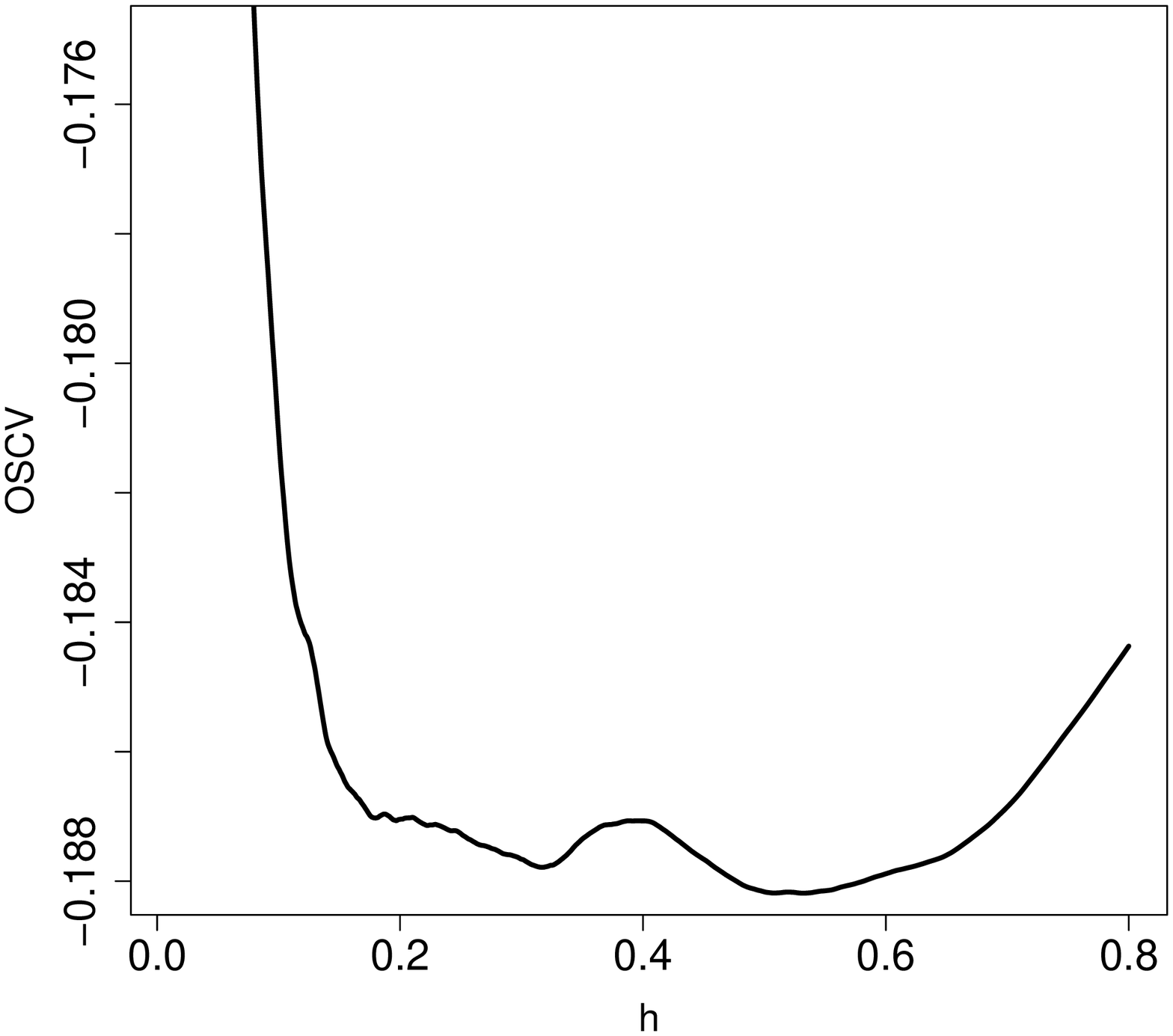,height=200pt}\\
\end{tabular}

\vspace{-0.5cm}
\caption{OSCV curves based on {\bf(a)} $L_G$ and {\bf(b)} $L_E$ for a random sample generated from $f^*$ at $n=500$.\label{fig:OSCV_realization_Epanechnikov}}
\end{center}
\end{figure}

Table~\ref{tab:measures_d4} shows the values of $\Delta_B$ and $\Delta_{ISE}$ from our simulations for three considered data-driven methods. The value of $\Delta_{ISE}=13.64$ for OSCV based on $C$ is much greater than $E_C=7.64$ from Table~\ref{tab:Constants}. This is explained by the fact that the OSCV method based on $L_G$ produces greatly variable bandwidths that tend to be inappropriately large even in the case when the nonsmooth constant $C^*$ is used. Figure~\ref{fig:OSCV_realization_Epanechnikov}~{\bf(a)} shows a typical $L_G$-based curve computed for a random sample generated from $f^*$ at $n=500$. The horizontal scale of the graph is changed such that the minimum is observed at the $C^*$-based OSCV bandwidth of 0.3179. The ISE-optimal bandwidth for these data is $\hat h_0=0.1673$.
\begin{table}
\begin{center}
\begin{tabular}{|c|c|c|c|}
\hline
Method&OSCV based on $C$&OSCV based on $C^*$&LSCV\\
\hline
$\Delta_B$&28.76&19.61&4.14\\
\hline
$\Delta_{ISE}$&13.64&10.00&6.57\\
\hline
\end{tabular}
\caption{Measures of performance computed from 1000 replications in the case of $f^*$ and $n=500$.\label{tab:measures_d4}}
\end{center}
\end{table}

\begin{figure}[h]
\begin{center}
\epsfig{file=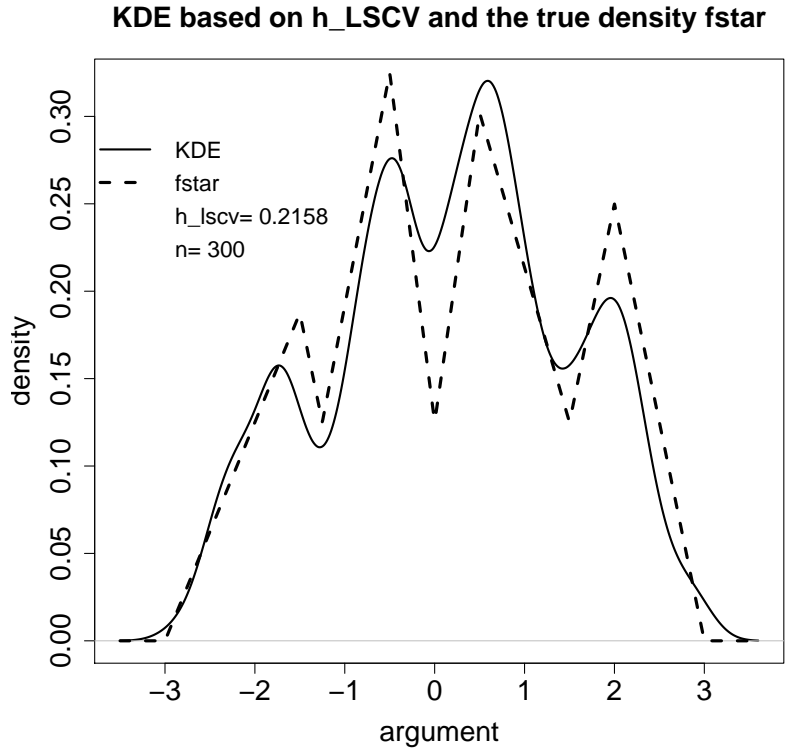,height=300pt}

\vspace{-0.75cm}
\caption{Density estimate based on the LSCV bandwidth for a realization generated from $f^*$ at $n=300$.\label{fig:fstar_ISEestimate_n300}}
\end{center}
\end{figure}

The LSCV method outperforms both OSCV versions in terms of the measures~\eqref{eq:delta_measures}. This is consistent with the aforementioned LSCV paradox since the density $f^*$ is relatively hard to estimate. Figure~\ref{fig:fstar_ISEestimate_n300} illustrates this by showing a density estimate based on the LSCV bandwidth for a random sample generated from $f^*$ at $n=300$. The ISE-optimal bandwidth for this data set is $\hat h_0=0.1884$.

The original implementation of the OSCV method of~\citet{OSCV:density} is based on $L_E$, the one-sided Epanechnikov kernel. 
Figure~\ref{fig:OSCV_realization_Epanechnikov}~{\bf(b)} shows the $L_E$-based OSCV curve computed for the same realization that yields the curve in Figure~\ref{fig:OSCV_realization_Epanechnikov}~{\bf(a)}. The horizontal scale of the graph in Figure~\ref{fig:OSCV_realization_Epanechnikov}~{\bf(b)} is changed such that its minimizer is to be plugged-in to the Gaussian density estimator without additional rescaling. However, the curve's minimizer of 0.4940 appears to be too large and results in an oversmoothed density estimate.

The curve in Figure~\ref{fig:OSCV_realization_Epanechnikov} {\bf(b)} is unacceptably wiggly. We generated many other samples from $f^*$ and found that the most of the corresponding $L_E$-based OSCV curves are inappropriately nonregular whereas the $L_G$-based OSCV curves are usually smooth and have one local minimum. This can be seen by experimenting with the R functions \verb"OSCV_Epan_dens" and \verb"OSCV_Gauss_dens" from the R package \verb"OSCV" (see~\citet{OSCV:package}).

The problem of producing insufficiently smooth OSCV curves  by $L_E$ persists in the smooth case, as the Example 2 illustrates. Roughness of the OSCV curves based on the Epanechnikov kernel was also noted in the regression context (see~\citet{Savchuk:FR_OSCV}).

\begin{figure}[h]
\begin{center}
\begin{tabular}{cc}
{\bf(a)}&{\bf(b)}\\
\vspace{-0.75cm}&\vspace{-0.75cm}\\
\epsfig{file=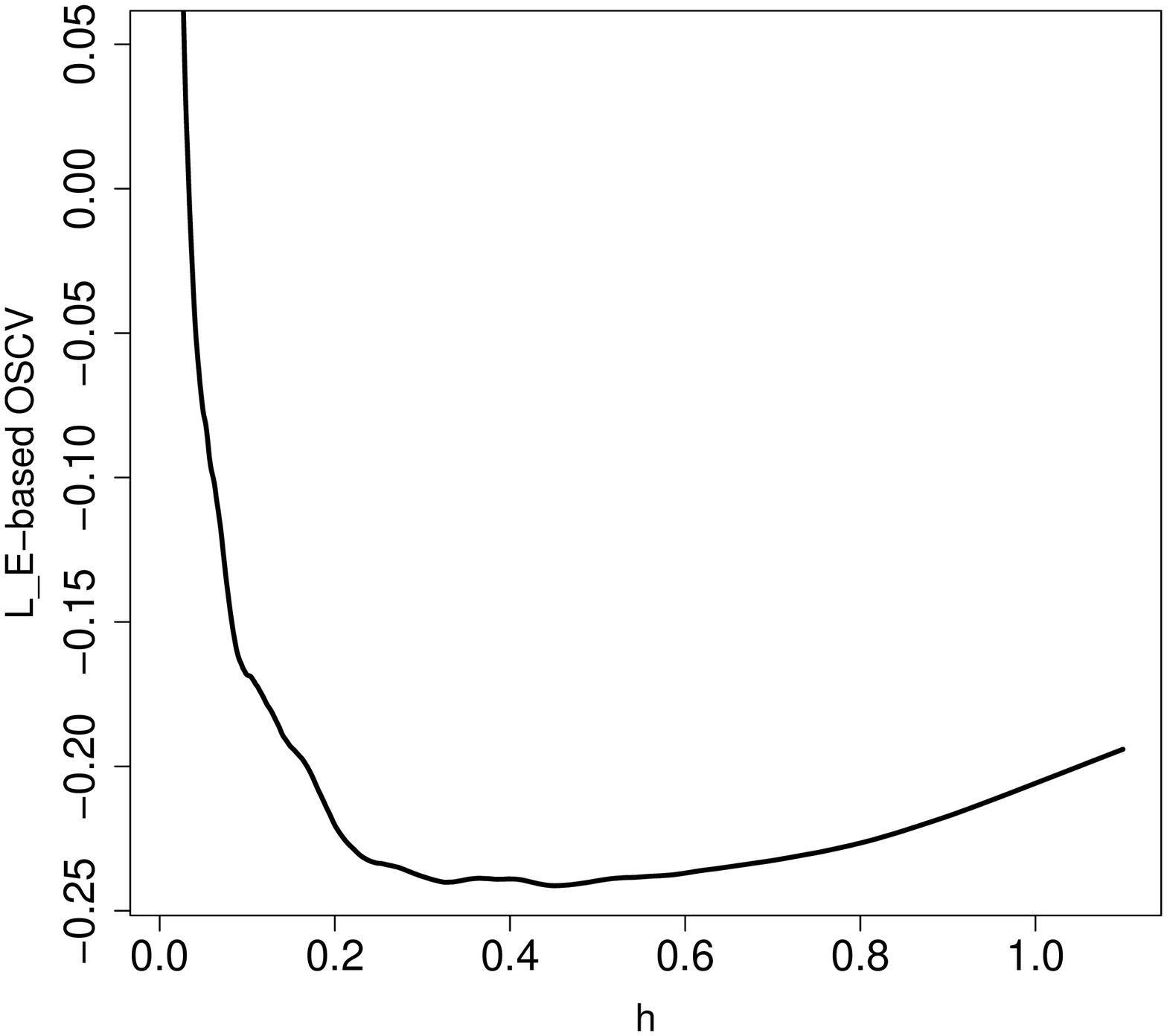,height=200pt}&\epsfig{file=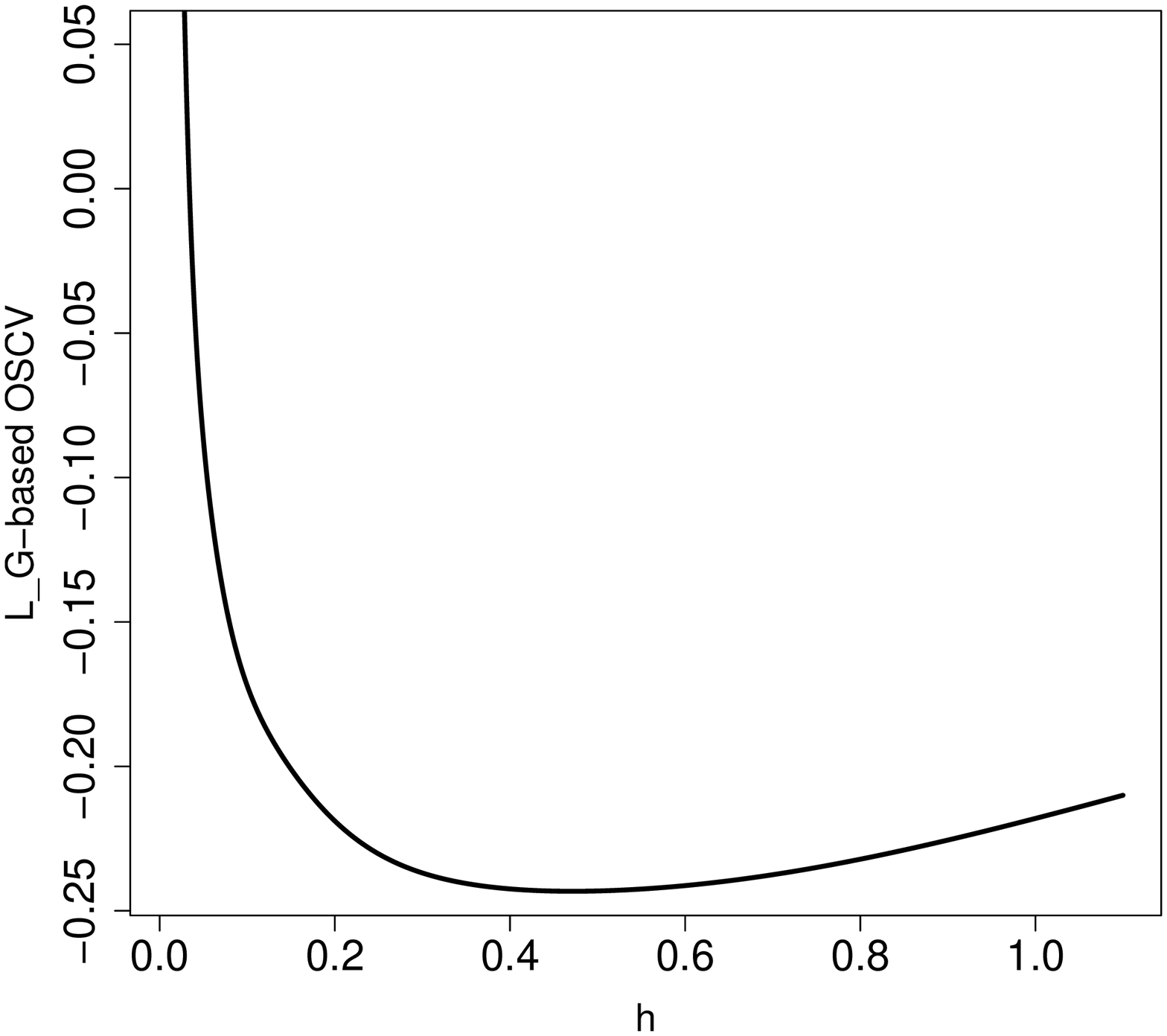,height=200pt}\\
\end{tabular}

\vspace{-0.5cm}
\caption{OSCV curves based on {\bf(a)} $L_E$ and {\bf(b)} $L_G$ for a random sample generated from the standard normal distribution at $n=100$.\label{fig:OSCV_realization_norm}}
\end{center}
\end{figure}

\vspace{0.25cm}
\noindent{\bf Example 2.} A random sample of size $n=100$ was generated from the standard normal density. Figure~\ref{fig:OSCV_realization_norm} shows the corresponding $L_G$\,-based and $L_E$-based OSCV curves. The curves' horizontal scales are adjusted such that their minimizers are to be used to compute the Gaussian density estimates without further rescaling. In fact, Figure~\ref{fig:OSCV_realization_norm} {\bf(a)} and {\bf(b)} shows the curves $\mbox{OSCV}_{L_G}(b/C_G)$ and $\mbox{OSCV}_{L_E}(b/C_E)$, where the rescaling constants $C_G$ and $C_E$ are obtained by using~\eqref{eq:C} with $K=\phi$ and $L$ corresponding to either $L_G$ or $L_E$, respectively.

The bandwidths selected by $L_E$ and $L_G$ are 0.4512 and 0.4714, correspondingly, whereas the ISE-optimal bandwidth $\hat h_0=0.4423$. The $L_E$-based OSCV curve is inappropriately wiggly. In particular, it has two local minima of about the same size located at 0.3268 and 0.4512. Just by luck the latter local minimum appears to be somewhat smaller.

For all other smooth densities that we considered in our numerous simulation experiments, the $L_G$-based OSCV curves were usually smoother than the corresponding $L_E$-based curves. Thus, $L_G$ appears to be a better candidate than $L_E$ for the OSCV method's implementation in the smooth case.

The problem of occasionally producing rough criterion curves by $L_E$  persists in the case of real data. This is illustrated by Example 3.

\vspace{0.25cm}
\noindent{\bf Example 3.} We used the famous data set of size $n=272$ on the eruption duration of the Old Faithful geyser in Yellowstone National Park that can be found in~\citet{Hardle:book}.
\begin{figure}[h]
\begin{center}
\begin{tabular}{cc}
{\bf(a)}&{\bf(b)}\\
\vspace{-0.55cm}&\vspace{-0.55cm}\\
\epsfig{file=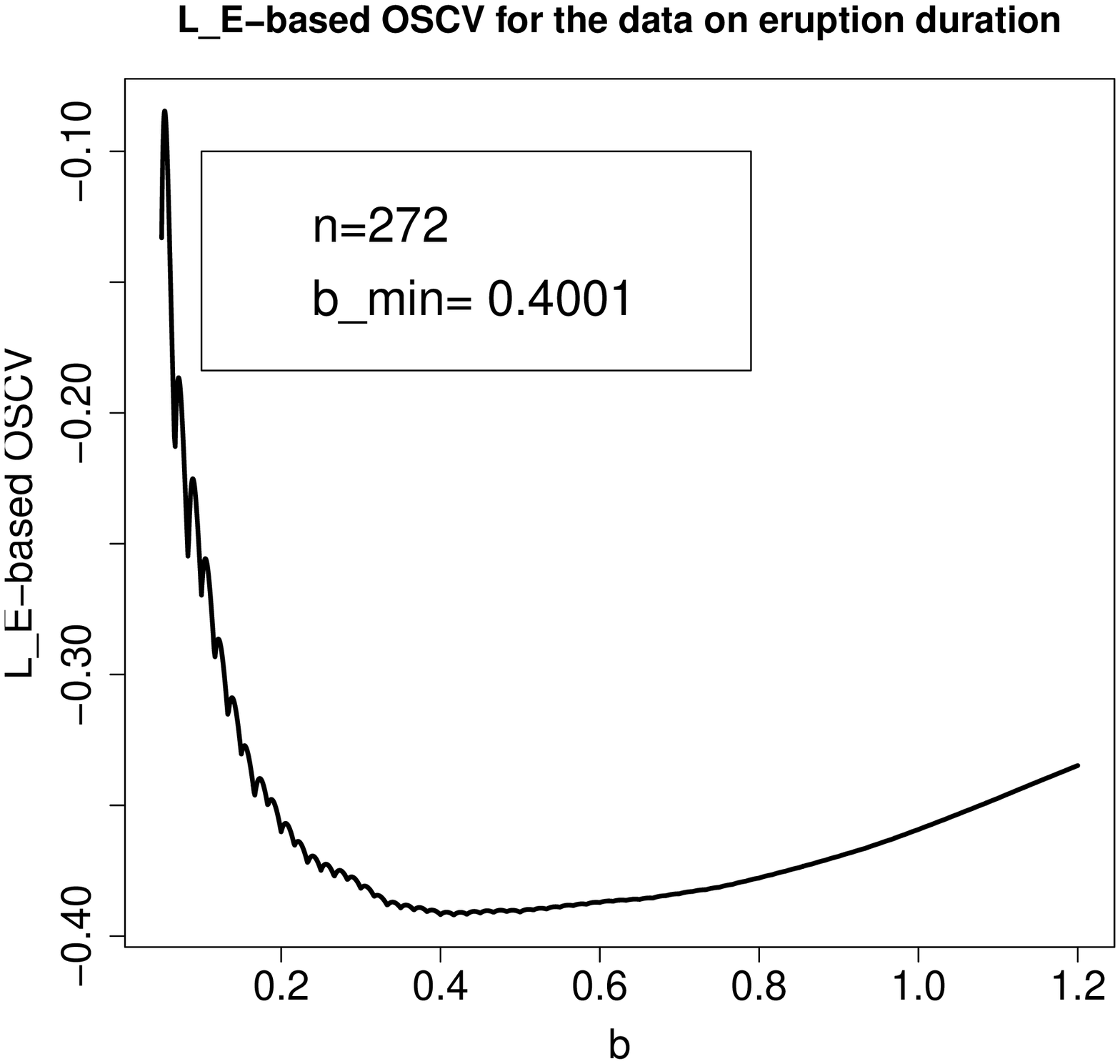,height=200pt}&\epsfig{file=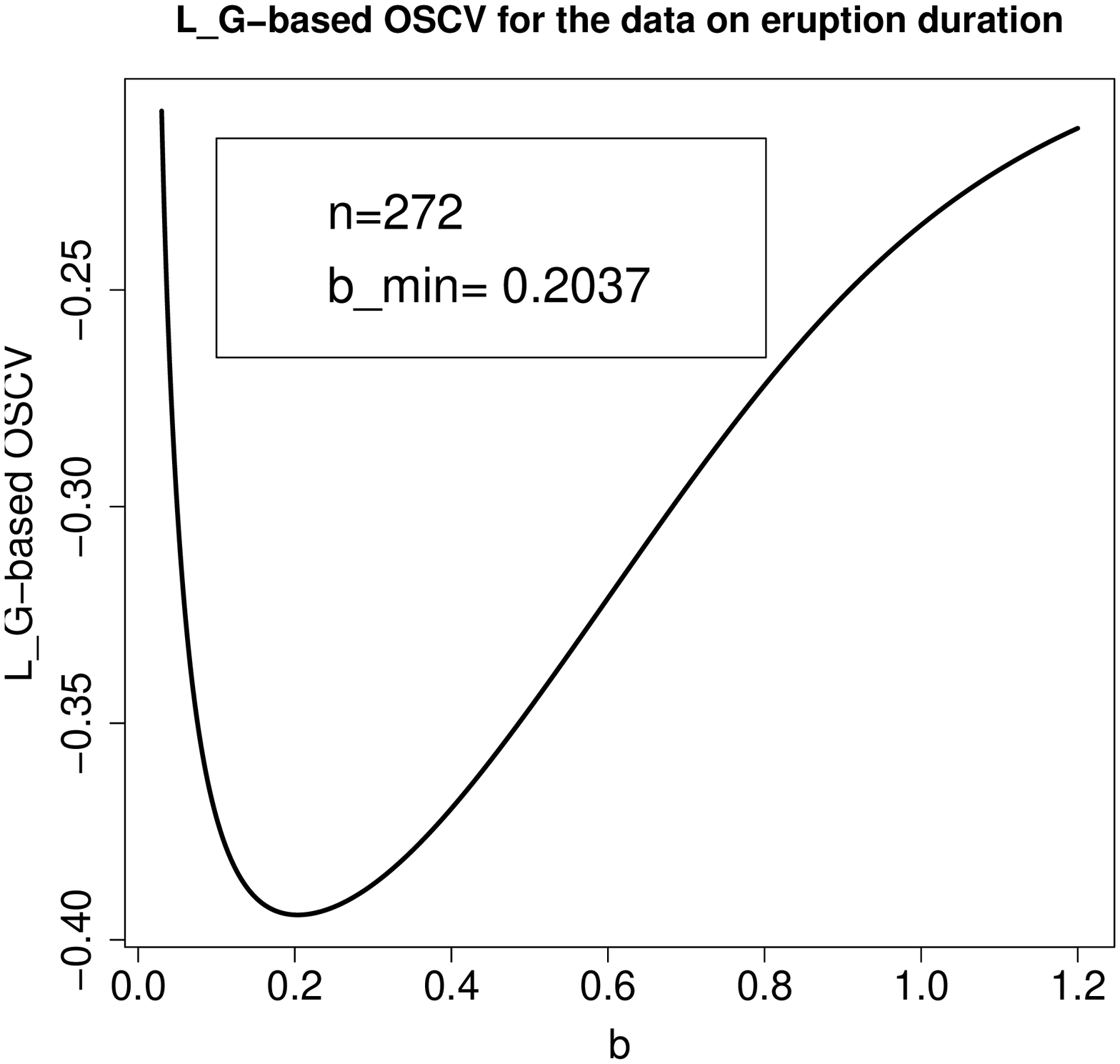,height=200pt}\\
\end{tabular}

\vspace{-0.5cm}
\caption{OSCV curves based on {\bf(a)} $L_E$ and {\bf(b)} $L_G$ for the data on eruption duration of the Old Faithful geyser. \label{fig:OSCV_geyser}}
\end{center}
\end{figure}
The corresponding $L_E$- and $L_G$-based OSCV curves are plotted in Figure~\ref{fig:OSCV_geyser}~{\bf(a)} and~{\bf(b)}, respectively. The horizontal scales of the graphs are not adjusted, so additional rescaling of the curves' minimizers is needed before they can be plugged into the Gaussian density estimator. The $L_E$~-based graph shown in Figure~\ref{fig:OSCV_geyser}~{\bf(a)} can be reproduced in R by using the code from the Examples section of the R package \verb"OSCV".

Obviously, the $L_E$-based curve for the eruption duration data is unacceptably wiggly even for the values of $b$ near the curve's minimizer. To the contrary, the $L_G$-based curve is perfectly regular. Assuming that the distribution of the eruption duration is smooth, the $L_G$-based OSCV bandwidth is found as $\hat h_{OSCV}=0.6168\cdot0.2037=0.1256$. For comparison, the LSCV and Sheather-Jones plug-in bandwidths are 0.1021 and 0.1395, respectively.

\section{Fully Robust OSCV for density functions\label{sec:FROSCV}}

The fully robust OSCV method proposed in the regression context (see~\citet{Savchuk:OSCVnonsmooth} and \citet{Savchuk:Corrigendum}) can be adapted for the density estimation setting. The underlying idea is that the OSCV function~\eqref{eq:OSCV} is computed based on a so-called {\it robust} one-sided kernel $L$ that has equal smooth and nonsmooth rescaling constants, $C$ and $C^*$, respectively. As a consequence, the OSCV bandwidth selected by a robust kernel is consistent for the MISE optimal bandwidth $h_0$ in both smooth and nonsmooth cases.

Since the density $f^*$ introduced in Example 1 is fairly nonsmooth, we used it for initial evaluation of performances of the robust and {\it almost} robust one-sided kernels that we managed to find. A kernel $L$ is called almost robust if it has $E_C<5\%$ (see~\citet{Savchuk:FR_OSCV}).
The same approach was previously used in the regression fully robust OSCV version where the performances of the robust and almost robust kernels were initially judged on the regression function $r_3$ that has six cusps. Thus, we concentrated on searching for a one-sided robust kernel $L$ that, at least, outperforms LSCV in the case of $f^*$ and produces reasonably smooth OSCV curves.
\begin{figure}[h]
\begin{center}
\begin{tabular}{cc}
{\bf(a)} Kernel with $(\alpha,\sigma)=(16.8954588,1.01)$&{\bf(b)} Kernel with $(\alpha,\sigma)=(0.4275,10)$\\

\vspace{-0.75cm}&\vspace{-0.75cm}\\

\epsfig{file=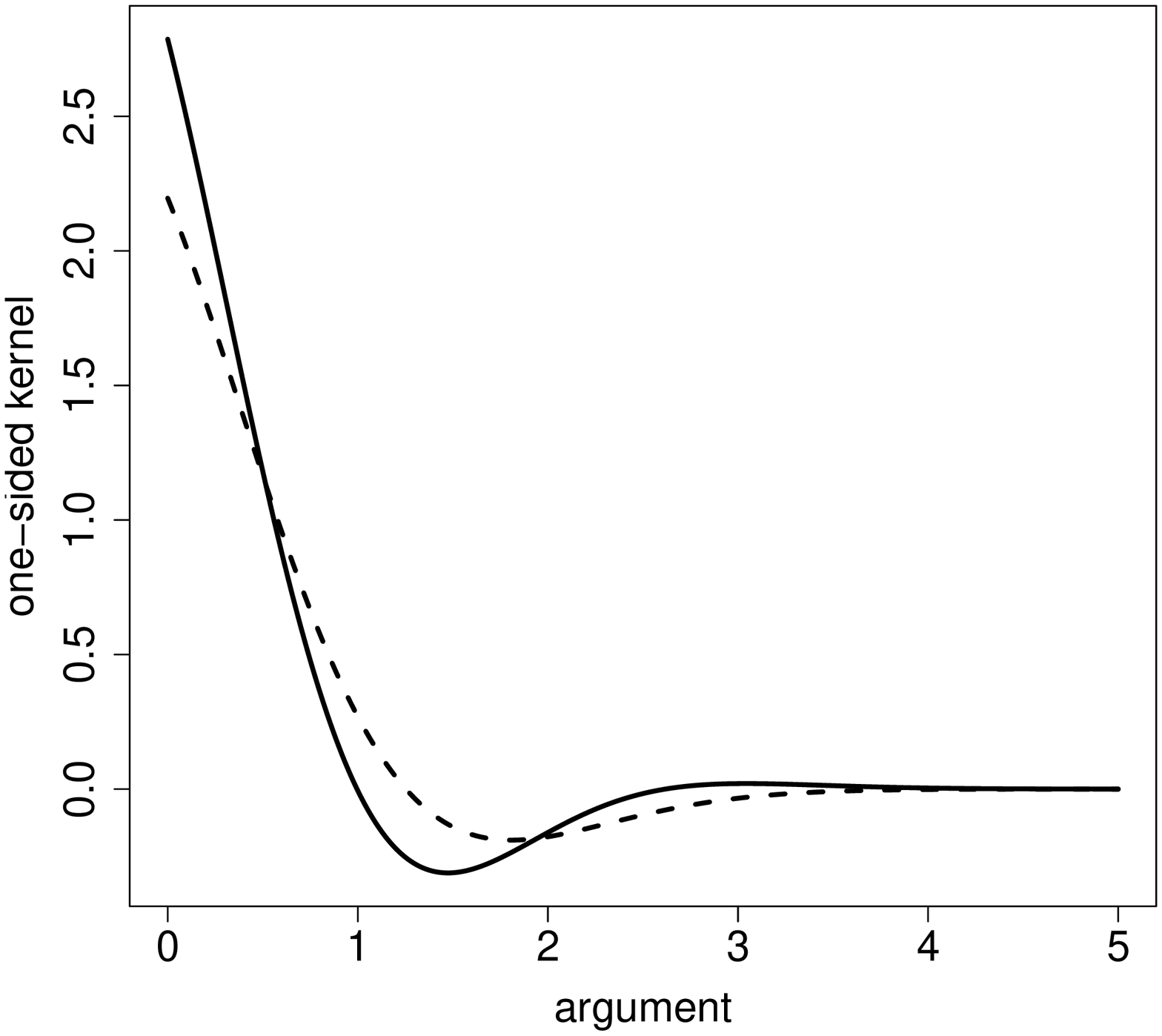,height=200pt}&\epsfig{file=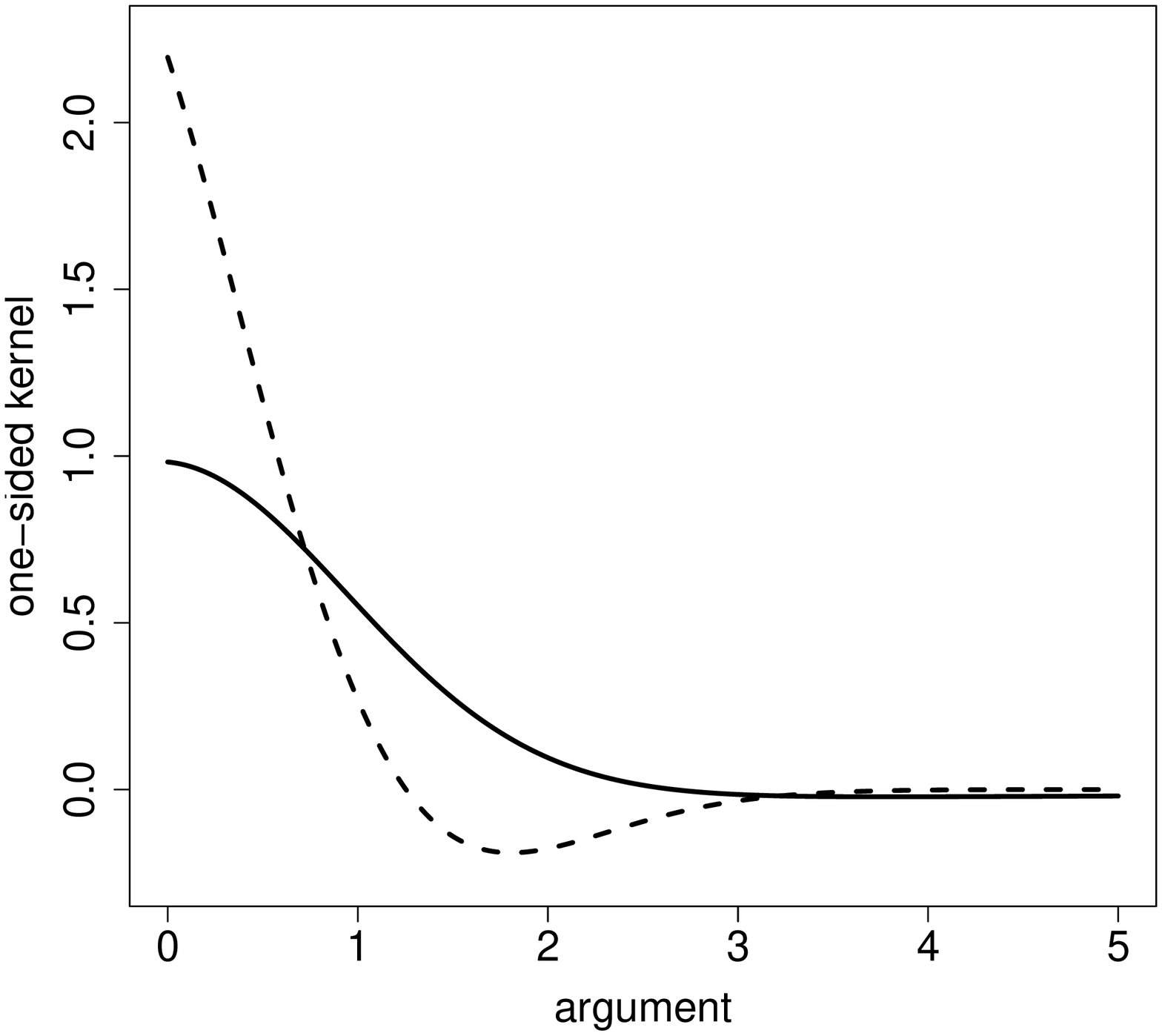,height=200pt}\\
{\bf(c)} Kernel with $(\alpha,\sigma)=(4,0.8)$&{\bf(d)} Kernel $L_1$\\

\vspace{-0.75cm}&\vspace{-0.75cm}\\

\epsfig{file=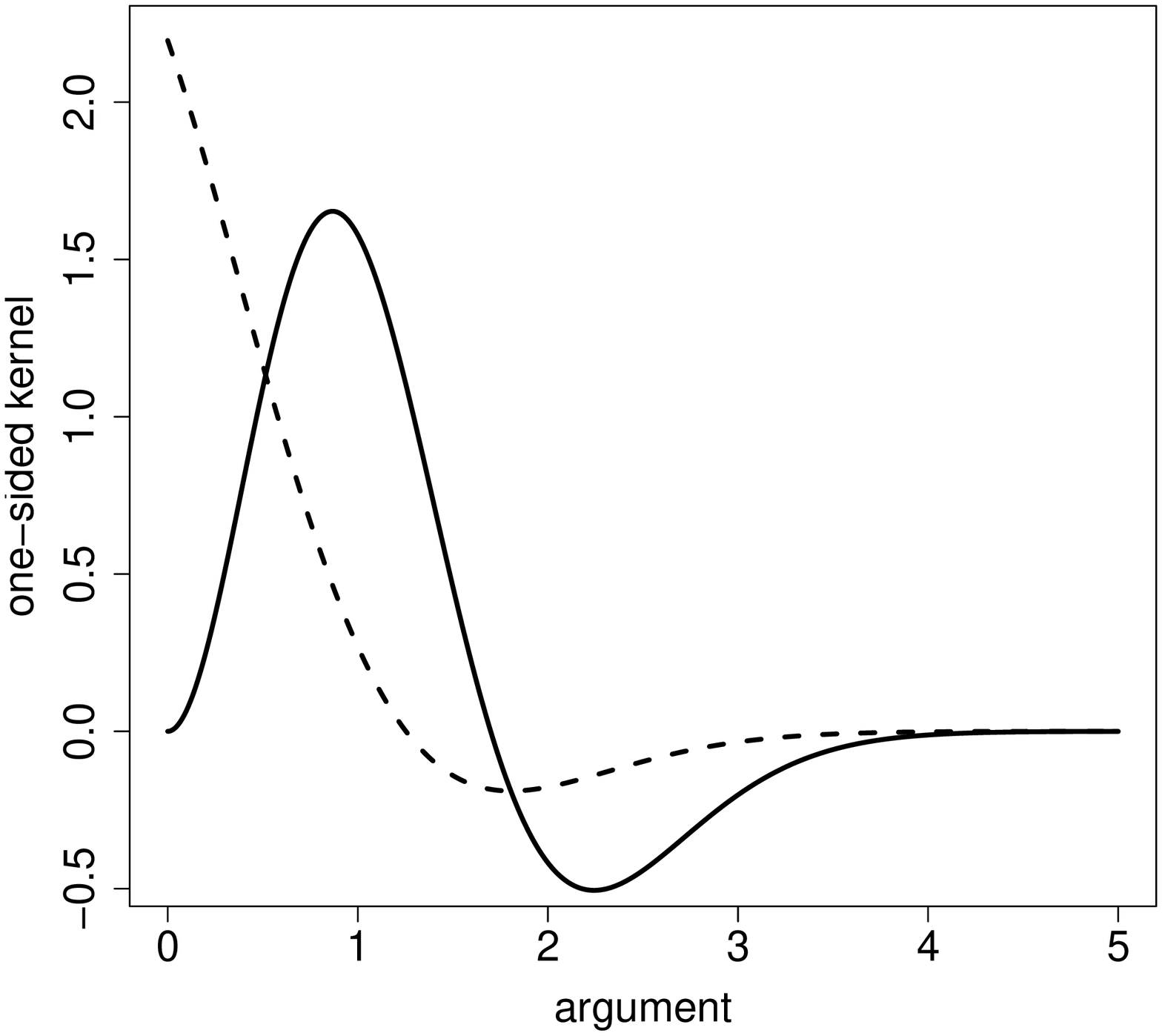,height=200pt}&\epsfig{file=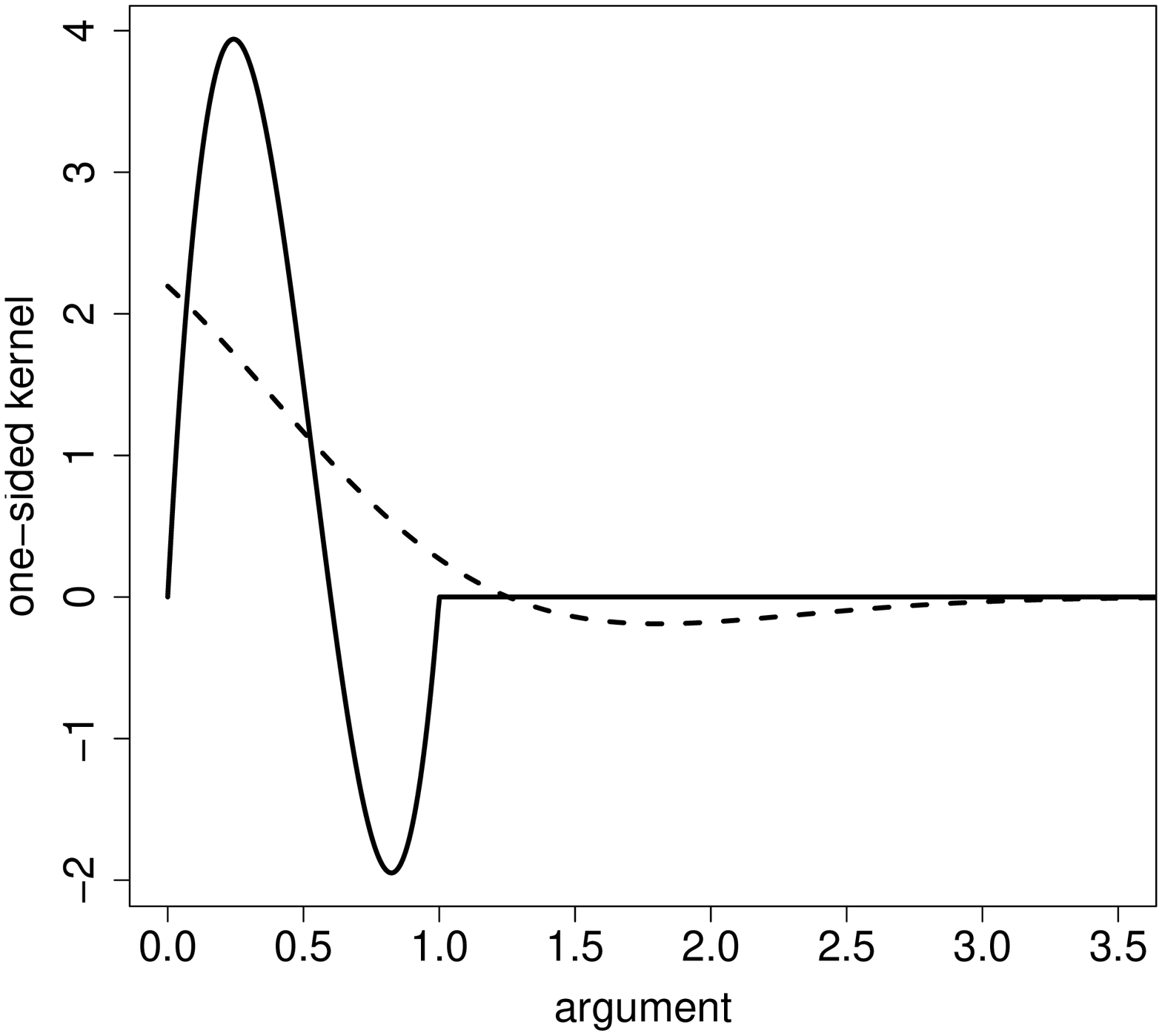,height=200pt}\\



\end{tabular}

\vspace{-0.5cm}
\caption{\label{fig:Robust_kernels}The solid curves show the one-sided kernels, whereas the dashed curves show $L_G$.}
\end{center}
\end{figure}

The robust kernel used in the fully robust OSCV implementation in the regression context (see ~\citet{Savchuk:FR_OSCV}) is also robust in the density estimation setting. This kernel is a member of the following family:
\begin{equation}
\label{eq:L_I}
L_I(u;\alpha,\sigma)=\frac{a+bu}{c}H_I(u)I_{[0,\infty)}(u),\qquad \alpha\in\mathbb{R},\ \sigma>0,
\end{equation}
where $H_I(u)=(1+\alpha)\phi(u)-\alpha\phi(u/\sigma)/\sigma$ is a two-sided counterpart of $L_I$, and
\[
\begin{array}{l}
\displaystyle{a=2\pi(1+\alpha-\alpha\sigma^2);}\\
\displaystyle{b=-2\sqrt{2\pi}(1+\alpha-\alpha\sigma);}\\
\displaystyle{c=\pi(1+\alpha-\alpha\sigma^2)-2(1+\alpha-\alpha\sigma)^2}.
\end{array}
\]
The kernel has $(\alpha,\sigma)=(16.8954588,1.01)$ and is plotted in Figure~\ref{fig:Robust_kernels}~{\bf(a)}. It is worth to mention that the one-sided Gaussian kernel $L_G$ is obtained from~\eqref{eq:L_I} by either setting $\alpha=0$ or $\sigma=1$.

Despite the fact that the kernel with $(\alpha,\sigma)=(16.8954588,1.01)$ works well in the regression context (see~\citet{Savchuk:FR_OSCV}), its performance in the density estimation framework is unsatisfactory. Our inspection of the $L_I$-based OSCV curves for a sequence of random samples generated from $f^*$ revealed that the considerable part of them has two local minima with the largest one being inappropriately large. However, the more serious problem with $L_I$ is that it frequently produces the OSCV curves that tend to $-\infty$ as $h\rightarrow 0$. \citet{Silverman:book} and~\citet{Chiu:rounding} argued that the LSCV method experiences this type of problem even for such frequently used kernels as Epanechnikov, quartic, and Gaussian.

Inappropriate performance of the kernel with $(\alpha,\sigma)=(16.8954588,1.01)$ on $f^*$ stimulated further search for the robust kernels. First of all, we inspected performances of two other
robust kernels mentioned in~\citet{Savchuk:FR_OSCV}. Both these kernels are members of the family~\eqref{eq:L_I} with $\sigma=10$ and the values of $\alpha$ equal to 0.4275 and 0.9821. One of the kernels (with $\alpha=0.4275$) is plotted in Figure~\ref{fig:Robust_kernels} {\bf(b)}. The other kernel (with $\alpha=0.9821$) has a similar shape and performance. Unfortunately, both kernels are found to produce highly variable and biased upwards bandwidth distributions, at least in the case of $f^*$.

In our next attempt, we considered another member of~\eqref{eq:L_I} with $(\alpha,\sigma)=(4,0.8)$ that is plotted in Figure~\ref{fig:Robust_kernels} {\bf(c)}. This kernel is almost robust with $E_C=1.17\%$. It has a quite different shape compared to the one-sided kernels considered so far, but, unfortunately, performs even worse than them. Indeed, for all inspected realizations from $f^*$ and the $N(0,1)$ density it produced the OSCV curves that tended to $-\infty$ as $h\rightarrow 0$. It is remarkable that the kernel with $(\alpha,\sigma)=(4,0.8)$ is equal to zero at the origin. This implies that $H_I$, the two-sided counterpart of this kernel, is nonnegative and bimodal (see~\citet{SavchukHartSheather:ICV}). According to~\citet{Savchuk:OSCVnonsmooth} and~\citet{Savchuk:FR_OSCV} such a kernel has potential for OSCV implementation in the regression context.

Further experimenting with the kernels from the family~\eqref{eq:L_I} is possible by using the R function \verb"OSCV_LI_dens" from the R package \verb"OSCV". Indeed, many other robust kernels can be found in~\eqref{eq:L_I}. It is entirely possible that there exists one that performs better than the kernels discussed above.

Three other almost robust kernels considered below are not the members of~\eqref{eq:L_I} but originate from the dissertation of~\citet{Yi:dissertation}. These kernels, denoted by $L_1$ $L_2$, and $L_3$, are defined below. Figure~\ref{fig:Robust_kernels} {\bf(d)} shows $L_1$. The graphs of the other two kernels are not included since they are fairly similar in shape to $L_1$.
\[
\begin{array}{l}
\displaystyle{L_1(u)=6u(1-u)(6-10u)I_{[0,1]}(u),}\\
\displaystyle{L_2(u)=30u^2(1-u)^2(8-14u)I_{[0,1]}(u),}\\
\displaystyle{L_3(u)=140u^3(1-u)^3(10-18u)I_{[0,1]}(u).}
\end{array}
\]
All of the above kernels have $|E_C|<0.3\%$. It appears that each of the kernels $L_1$, $L_2$, and $L_3$ produces quite wiggly OSCV curves for random samples generated from $f^*$ and the $N(0,1)$ density.

The Figure~\ref{fig:Robust_kernels} shows variety of robust and almost robust one-sided kernels of different shapes, but none of them performs satisfactory in the case of $f^*$. Thus, finding a kernel that improves practical performance of the method in the nonsmooth case appears to be an open challenging problem.

\section{Summary}

The OSCV method for smooth density functions is proposed by~\citet{OSCV:density}. In this article we extend the OSCV methodology to the case of nonsmooth densities. We also introduce the fully robust OSCV version that produces consistent bandwidths in both smooth and nonsmooth cases.

The proposed OSCV modifications, essentially, use different one-sided kernels to select the bandwidths by the cross-validation method. The selected bandwidths are then rescaled and plugged-in to the Gaussian density estimator.
The nonsmooth density $f^*$ plotted in Figure~\ref{fig:f_nonsmooth}~{\bf(a)} is used for discrimination of the proposed OSCV extensions.

We found and investigated many robust and almost robust kernel candidates for the fully robust OSCV implementation. Some of them are shown in Figure~\ref{fig:Robust_kernels}. None of these kernels performs satisfactory in the case of $f^*$. Moreover, the nonsmooth version of OSCV based on the one-sided Gaussian kernel $L_G$ performs worse that the ordinary LSCV method in the case of $f^*$. Thus, practical implementations of the proposed theoretical extensions remain open to further research efforts.

The main problems experienced by the majority of the considered robust and almost robust one-sided kernels were selecting too variable bandwidths and/or producing multiple local minima in the OSCV curves. Similar difficulties were faced when implementing the ICV method (see ~\citet{SavchukHartSheather:ICV}, \citet{SavchukHartSheather:empiricalICV}, and~\citet{ICV:package}). This indicates that some nontraditional negative-valued cross-validation kernels may substantially improve the asymptotic properties of the cross-validation method while introducing challenging problems with their practical use.

The current implementation of the OSCV method in the smooth case of~\citet{OSCV:density} is based on the one-sided Epanechnikov kernel $L_E$. It appears that $L_E$ produces inappropriately wiggly OSCV curves in the case of $f^*$ (see Figure~\ref{fig:OSCV_realization_Epanechnikov} {\bf(b)}). Moreover, $L_E$ frequently yields insufficiently smooth OSCV curves even in the case of the standard normal density (see Figure~\ref{fig:OSCV_realization_norm} {\bf(a)}). The problem of $L_E$ occasionally producing rough criterion curves persists for real data sets (see Figure~\ref{fig:OSCV_geyser} {\bf(a)}). On the other hand, we empirically found that for variety of smooth and nonsmooth densities and different sample sizes, the one-sided Gaussian kernel $L_G$ usually produces smooth OSCV curves with one local minimum (see Figures~\ref{fig:OSCV_realization_Epanechnikov} {\bf(a)}, \ref{fig:OSCV_realization_norm} {\bf(b)} and~\ref{fig:OSCV_geyser} {\bf(b)} for illustration). This indicates that $L_G$ might be, potentially, superior than $L_E$ for practical implementation of the OSCV method in the smooth case. This matter, however, requires further investigation that is out of scope of this article that is mainly devoted to extending OSCV to the case of nonsmooth densities.

The almost robust one-sided kernel shown in Figure~\ref{fig:Robust_kernels}~{\bf(c)} with $(\alpha,\sigma)=(4,0.8)$ has a nonnegative two-sided counterpart $H_I$. It then follows from the conclusions of~\citet{Savchuk:OSCVnonsmooth} and~\citet{Savchuk:FR_OSCV} that this kennel might, potentially, lead to successful implementation of the fully robust OSCV method in the regression context.

This article and the recent publication of~\citet{Savchuk:FR_OSCV} are supported by the R package \verb"OSCV" that can be used for reproducing the presented results and allows for further experimenting in attempts of improving the OSCV method's practical implementation for smooth and nonsmooth density and regression functions.

\bibliographystyle{abbrvnat}

\section{Appendix}

\noindent{\bf Notation.}

\noindent For an arbitrary function $g$, define the following functionals:
\begin{equation*}
\begin{array}{l}
\displaystyle{\mu_2(g)=\int_{-\infty}^\infty u^2 g(u)\,du},\\[0.3cm]
\displaystyle{R(g)=\int_{-\infty}^\infty g^2(u)\,du},\\[0.3cm]
\displaystyle{D_g(z)=\int_{-\infty}^z g(u)\,du},\\[0.3cm]
\displaystyle{G_g(z)=\int_{-\infty}^z ug(u)\,du,\qquad z\in\mathbb R}.
\end{array}
\end{equation*}
Based on $D_g$ and $G_g$ we define
\begin{equation}
\label{eq:B}
B(g)=\int_0^\infty\left\{z\bigl(1-D_g(z)\bigr)+G_g(z)\right\}^2\,dz+\int_0^\infty\left\{zD_g(-z)+G_g(-z)\right\}^2\,dz.
\end{equation}

\end{document}